\documentclass[10pt,twocolumn]{article}

\usepackage[letterpaper,margin=0.72in,columnsep=0.25in]{geometry}
\usepackage{microtype}
\usepackage{graphicx}
\usepackage{booktabs}
\usepackage{array}
\usepackage{amsmath}
\usepackage{amssymb}
\usepackage{xcolor}
\usepackage[hidelinks]{hyperref}
\hypersetup{
  pdftitle={The Acknowledgment Point Is the System: Durable Policy-Decision Receipts for AI Audit Evidence},
  pdfauthor={Neeraj Kumar Singh Beshane},
  pdfsubject={Independent research on durable and verifiable AI policy-decision evidence},
  pdfkeywords={AI accountability, audit evidence, durable logging, Merkle tree, policy enforcement}
}
\usepackage{url}
\usepackage{enumitem}
\setlist{nosep,leftmargin=*}
\newcommand{\system}{RuntimeGuard-AI V2}
\newcommand{\sha}{SHA-256}

\def\RGSourceDigestShort{c78c0c2219e6}

\def\RGThreads{4}
\def\RGPromptBytes{2,048}
\def\RGPolicyMedian{0.938}
\def\RGPolicyTail{3.146}
\def\RGPolicyThroughput{1,717,597.126}

\def\RGBufferedMedian{141.875}
\def\RGBufferedTail{238.145}
\def\RGBufferedThroughput{27,192.808}

\def\RGDataMedian{16,018.646}
\def\RGDataTail{33,963.302}
\def\RGDataThroughput{240.123}

\def\RGFullMedian{16,009.136}
\def\RGFullTail{29,700.826}
\def\RGFullThroughput{244.416}

\def\RGEpochRecords{100,000}
\def\RGEpochSealUS{96,954.958}
\def\RGEpochProofUS{3.979}
\def\RGEpochSignatureUS{25.75}
\def\RGRecoveryRecords{100,000}
\def\RGRecoveryOpenMS{665.483}
\def\RGRecoveryReadSortMS{295.487}

\title{The Acknowledgment Point Is the System:\\Durable Policy-Decision Receipts for AI Audit Evidence}
\author{Neeraj Kumar Singh Beshane\\Independent Researcher, Fremont, California}
\date{}

\begin{document}
\maketitle

\begin{abstract}
An AI audit record is useful only if its durability and trust boundary are explicit. Returning a guarded decision before any durable write minimizes latency, but it cannot guarantee that evidence survives an immediate crash. We rebuild RuntimeGuard-AI around this constraint. The resulting research prototype binds each deterministic policy decision to the exact policy source, commits a privacy-minimizing record at a caller-selected synchronization boundary, and returns an Ed25519-signed receipt that states whether that boundary completed. After restart, the engine validates framed records, manifests, shard placement, sequence continuity, and replay identity. A separate attestation path groups committed records into chained, signed Merkle epochs that an auditor verifies with an externally obtained key. On an Apple M4 Pro at four worker threads and 2,048-byte prompts, buffered signed evidence reaches 27,193 requests/s with 141.9~$\mu$s median latency. Per-record data and full synchronization reduce throughput to approximately 242 requests/s and raise median latency to 16.0~ms. Sealing a 100,000-record signed epoch takes 97.0~ms. The result is a measured durability--latency trade-off, not a ``free'' asynchronous audit path. The prototype does not prove model execution, prevent a compromised signer from forking history, or establish legal conformity.
\end{abstract}

\section{Introduction}
The hard part of runtime accountability is not hashing a log. It is deciding when a system may truthfully tell a caller that evidence exists.

Consider a guarded AI request. A policy evaluator can return immediately and schedule logging in the background. This makes the fast path attractive, but creates an unavoidable interval in which the action has been released and its evidence exists only in volatile state. A crash in that interval erases the record. Conversely, waiting for storage synchronization makes the durability claim meaningful but puts storage latency on the request path. No queue, detached task, or cryptographic proof removes this trade-off.

The version-of-record RuntimeGuard-AI paper described an asynchronous zero-knowledge attestation architecture and reported performance claims that the released prototype did not reproduce~\cite{runtimeguard_vor}. This successor treats that mismatch as a design failure rather than a documentation defect. We remove the unrelated proof circuit and rebuild the smallest implementation whose claims can be tested end to end.

\system{} centers the protocol on a signed \emph{commit receipt}. The receipt binds the request identifier and commitment, committed record, sequence number, durability bit, and verification key. In data- and full-synchronization modes, the engine returns the receipt only after the configured host synchronization call succeeds. Buffered mode remains available, but its receipt explicitly states that the record is not durable. This distinction turns durability from an implied property into a machine-checkable part of the interface.

The work makes four contributions:
\begin{enumerate}
  \item a source-bound policy and record format with deterministic, domain-separated commitments;
  \item a single-writer commit protocol with explicit synchronization semantics, signed receipts, idempotent replay, fail-stopped write errors, and strict restart validation;
  \item chained Ed25519-signed Merkle epochs and an end-to-end verifier from retained receipt to trusted epoch inclusion; and
  \item a preregistered, source-hashed component benchmark that reports the cost of each durability mode rather than collapsing them into one ``low-overhead'' claim.
\end{enumerate}

These are systems-integration and measurement contributions, not new cryptography. The paper therefore evaluates implementation fidelity and cost, and states where stronger transparency, trusted execution, or distributed protocols would be required.

\section{Problem and Guarantee Boundary}
\subsection{The commit contradiction}
Let $D$ be a policy decision and $A$ the event that the system acknowledges it to the caller. Let $W$ be completion of a write to volatile operating-system state and $S$ completion of the selected host synchronization boundary. If $A$ precedes both $W$ and $S$, an immediate process or power failure can occur after acknowledgment but before persistence. Therefore, a system cannot simultaneously guarantee crash-surviving evidence and acknowledge before any durability operation.

\system{} exposes three modes:
\begin{itemize}
  \item \textbf{buffered}: append a framed record and return a signed receipt with \texttt{durable=false};
  \item \textbf{data sync}: call \texttt{sync\_data} before returning \texttt{durable=true}; and
  \item \textbf{full sync}: call \texttt{sync\_all} before returning \texttt{durable=true}.
\end{itemize}
The last two statements are conditional on the operating system, filesystem, storage controller, and device honoring their documented semantics. They are not claims of remote replication or immunity to rollback.

\subsection{Threat model}
We consider interruption during append, incomplete tail writes, corruption of complete frames, replayed requests, conflicting reuse of a request identifier, record gaps or duplication, policy or record tampering, invalid Merkle proofs, modified epoch statements, and attacker-generated signatures under an untrusted key.

We trust the loaded policy source, the executing binary, host storage semantics, uncompromised receipt and epoch keys, and verification keys obtained independently by the auditor. A retained signed receipt lets a client challenge an operator to produce the matching committed record and inclusion proof.

We do not defend against a root adversary that replaces code, keys, logs, and verifier configuration together. We also exclude rollback to an older internally valid snapshot, signer forks without an external observer, bypass of the guarded call site, key revocation, and distributed exactly-once execution. Hash commitments bind data but do not make low-entropy fields confidential.

\section{Protocol}
\begin{figure*}[t]
  \centering
  \includegraphics[width=\textwidth]{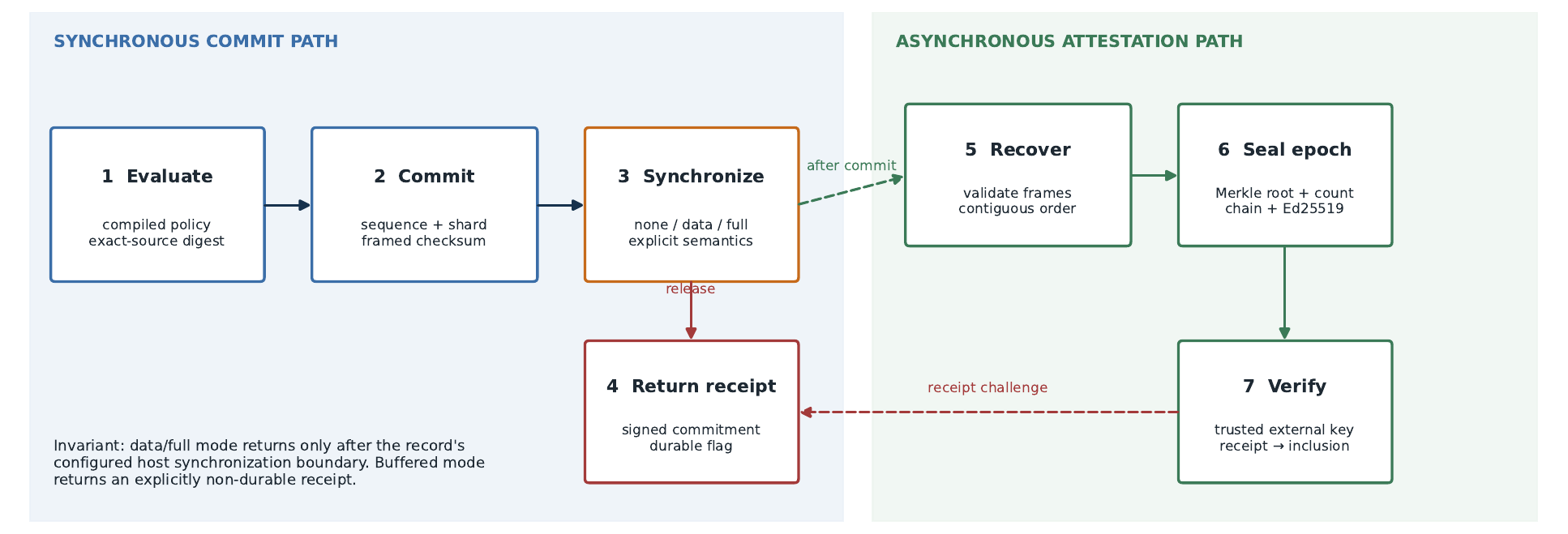}
  \caption{RuntimeGuard-AI V2 lifecycle. The synchronous path returns only after its selected host synchronization boundary. Epoch construction and audit verification operate on committed records outside that path.}
  \label{fig:lifecycle}
\end{figure*}

Figure~\ref{fig:lifecycle} separates the synchronous commit path from asynchronous epoch construction. The separation is about expensive batch attestation, not about eliminating storage cost from durable acknowledgment.

\subsection{Policy and request binding}
A compiled policy contains canonical source bytes and a descriptor $(id, version, digest)$, where $digest=\sha(source)$. The evaluator consumes those same source bytes; callers cannot attach an arbitrary descriptor to unrelated logic. A request commitment is
\[
H_R = \sha(\texttt{request-domain} \parallel \operatorname{enc}(R)),
\]
where \(\operatorname{enc}\) uses fixed-width lengths and a fixed field order. The compliance record stores commitments and selected metadata, not the raw prompt or input payload.

\subsection{Commit state machine}
One engine instance holds an operating-system-backed exclusive writer lease on the evidence directory. A second process cannot open the same directory for writing until the lease is released. Within that writer, one commit mutex serializes sequence assignment and append. The serialized point is deliberate: a later record cannot overtake a sequence whose append has not completed.

For a new request, the engine evaluates the compiled policy, allocates the next global sequence, selects shard $sequence \bmod K$, constructs a protocol-stable record commitment, appends one checksummed frame, and applies the selected synchronization operation. It then signs the receipt. The JSON frame is a storage encoding; fixed binary encodings define cryptographic commitments.

\begin{figure*}[t]
  \centering
  \includegraphics[width=0.94\textwidth]{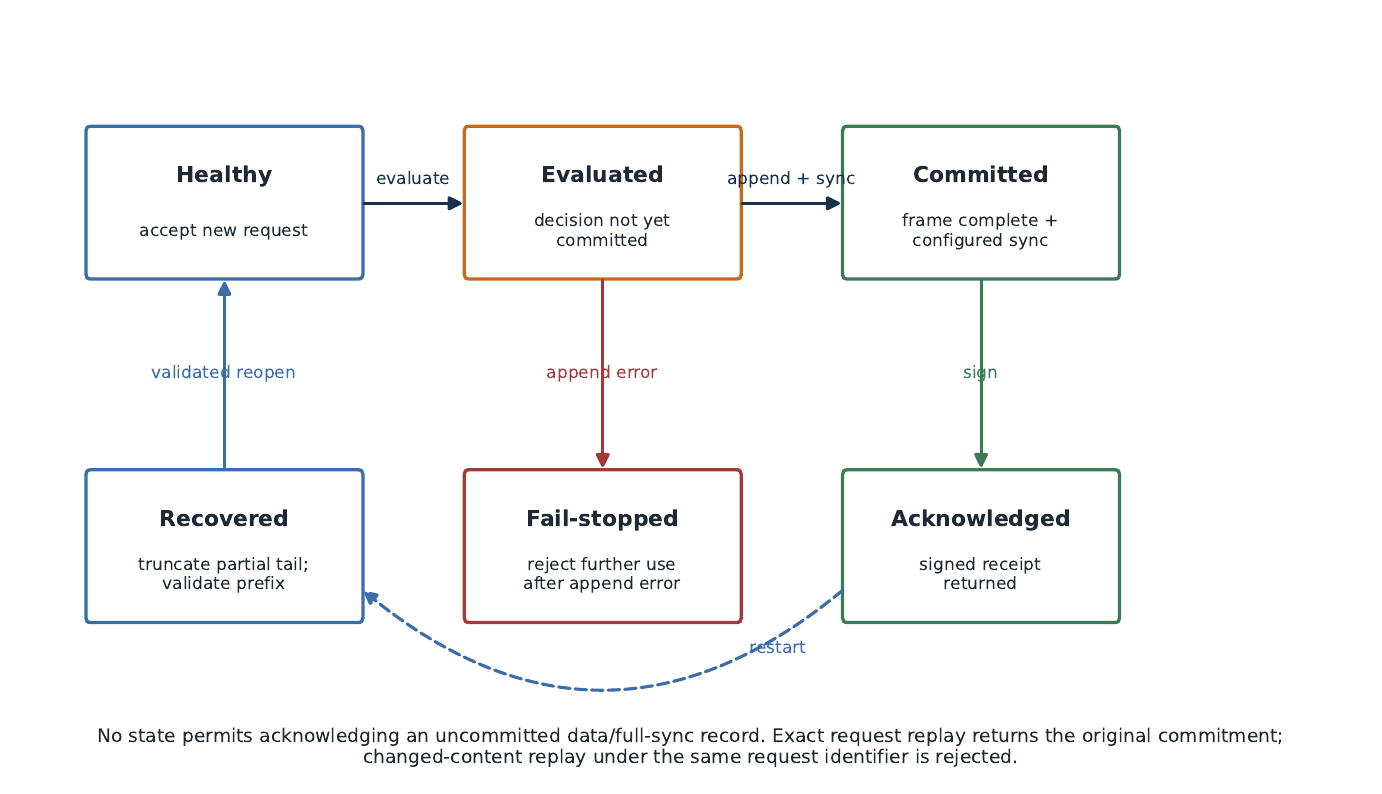}
  \caption{Commit and recovery states. No data- or full-sync request reaches \emph{Acknowledged} before a complete frame and its configured synchronization operation.}
  \label{fig:state}
\end{figure*}

If append or synchronization fails, the engine enters a fail-stopped state. It does not consume the sequence and continue, because doing so would create a gap that later recovery must either hide or reject. Figure~\ref{fig:state} summarizes the transitions.

Exact replay of the same request identifier and commitment returns the original decision and a deterministically reconstructed receipt without a second record. Reuse of the identifier with different committed content is rejected. The lookup is repeated after acquiring the commit mutex, so concurrent identical retries still append once.

\subsection{Recovery}
Each shard uses an \texttt{RGL2} frame containing magic and version bytes, payload length, JSON payload, and an unkeyed frame checksum. The checksum detects accidental or unsophisticated corruption; it is not an adversarial authenticity mechanism because an operator can recompute it.

On open, the engine acquires the writer lease and validates a manifest that pins format, shard count, synchronization mode, policy digest, and receipt verification key. It truncates only an incomplete final frame and rejects corruption in any complete frame. Across recovered records it verifies record commitments, shard placement, unique request identifiers, and the exact global sequence prefix $0,\ldots,N-1$. A validly checksummed gap is therefore rejected rather than silently accepted.

\subsection{Signed epochs and audit}
The attestation path sorts a non-empty, single-policy, contiguous record range and builds a \sha{} Merkle tree. Leaf and internal-node hashes use different domains. Odd levels duplicate the last node, and each proof is verified against the signed logical leaf count to avoid structural ambiguity.

An epoch statement binds the root, logical count, first and last sequence, first and last evaluation time, policy descriptor, predecessor statement hash, and signer key identifier. The statement is signed with Ed25519. Verification requires an externally supplied trusted key; the key serialized beside a signature is never accepted as its own trust root. Successor verification checks both signatures, the predecessor statement hash, and sequence adjacency.

An auditor verifies a retained receipt, resolves its record commitment, verifies the epoch signature and policy equality, derives the proof index from the sequence range, and verifies Merkle inclusion. Without an external witness, this proves integrity of an observed epoch chain, not global absence of forks or omitted pre-seal requests.

\section{Implementation}
The prototype is a Rust workspace with separate inline and attestor crates. The inline crate owns policy compilation, deterministic commitments, sharded framed logs, the writer lease, receipts, restart validation, and inline/recovery benchmarks. The attestor crate owns Merkle trees, signed epoch statements, chain verification, inclusion verification, and the epoch benchmark.

Security-property tests cover incomplete-tail truncation, complete-frame corruption, manifest mismatch, zero-shard rejection, sequence gaps, changed-content replay, concurrent exact replay, concurrent durable commits, writer-lease exclusion, source-bound policies, trusted receipt keys, odd-sized Merkle trees, logical-size-bound proofs, attacker signing keys, statement mutation, epoch gaps, chain continuity, and receipt-to-epoch inclusion.

The implementation deliberately omits the version-of-record's dummy Groth16 circuit. A signature proves that a trusted key signed a statement; it does not prove that model inference or arbitrary policy code executed correctly. Such a claim would require trusted measurement of the evaluator or a proof relation that encodes it.

\section{Experimental Method}
We ask four questions: (RQ1) what cost signed evidence and each synchronization boundary add to policy evaluation; (RQ2) how worker count and prompt size affect the closed-loop component paths; (RQ3) how epoch construction, proof generation, inclusion verification, and signature verification scale; and (RQ4) how log opening and recovery scale with retained records.

The full inline matrix crosses four modes (policy only, buffered evidence, data sync, full sync), three worker counts (1, 4, 8), three prompt sizes (128, 2,048, 16,384 bytes), and 20 repetitions. Each repetition uses 200 warm-up and 1,000 measured requests. Condition order is randomized with a fixed seed. Epochs use 100 to 100,000 records, and recovery uses 1,000 to 100,000 records; both use five warm-ups and 30 measured repetitions per scale.

We record per-request latency and whole-condition elapsed time. Within each repetition we compute p50, p95, and p99 latency and wall-clock throughput. Across independent repetitions we report medians and deterministic 10,000-resample percentile-bootstrap 95\% confidence intervals. Evidence overhead is paired with the policy-only repetition sharing worker count and prompt size.

The runner refuses output-directory reuse, records redacted hardware and exact tool versions, hashes every source file that can affect behavior or analysis, executes formatting, warnings-as-errors linting, all-target tests, and dependency audit before measurement, and hashes every raw artifact. The analyzer verifies those hashes before producing summaries or figures. Source drift invalidates the corpus.

\section{Results}
The canonical run completed all 720 inline conditions, four epoch scales, and three recovery scales. Its end-of-run source digest (\texttt{\RGSourceDigestShort}) matched the starting digest; all raw and derived artifact hashes verified. Table~\ref{tab:inline} reports the preregistered four-thread, 2,048-byte condition. Each estimate is the median over 20 independent randomized repetitions.

\begin{table}[t]
\centering
\small
\caption{Inline component results at \RGThreads{} threads and \RGPromptBytes-byte prompts. Latency is in microseconds; throughput is requests/s.}
\label{tab:inline}
\resizebox{\columnwidth}{!}{%
\begin{tabular}{lrrr}
\toprule
Mode & p50 & p99 & Throughput \\
\midrule
Policy only & \RGPolicyMedian & \RGPolicyTail & \RGPolicyThroughput \\
Buffered evidence & \RGBufferedMedian & \RGBufferedTail & \RGBufferedThroughput \\
Data sync & \RGDataMedian & \RGDataTail & \RGDataThroughput \\
Full sync & \RGFullMedian & \RGFullTail & \RGFullThroughput \\
\bottomrule
\end{tabular}
}
\end{table}

\paragraph{RQ1: durability dominates the cost.}
Policy evaluation alone is sub-microsecond at the selected condition. Constructing, appending, and signing buffered evidence raises median latency to \RGBufferedMedian~$\mu$s and sustains \RGBufferedThroughput{} requests/s. Data and full synchronization raise median latency to \RGDataMedian{} and \RGFullMedian~$\mu$s, respectively, while throughput falls to \RGDataThroughput{} and \RGFullThroughput{} requests/s. Relative to buffered evidence, per-record synchronization is approximately 112$--$113$\times$ slower in throughput and 112.9$\times$ higher in median latency. Data and full sync are numerically similar on this host; the experiment does not establish equivalence or generalize the comparison to other filesystems or devices.

\paragraph{RQ2: serialization saturates rather than scales.}
At 2,048-byte prompts, buffered throughput is 30,884, 27,193, and 30,094 requests/s at 1, 4, and 8 threads. Per-record synchronized throughput remains near 243 requests/s across worker counts. Because one commit lock preserves global order, additional closed-loop workers wait: synchronized median latency grows from approximately 4~ms at one thread to 16~ms at four and 32~ms at eight. Sharding distributes files but does not create parallel sequence authorities. This is a measured design cost, not a scalability result.

\begin{figure*}[t]
  \centering
  \includegraphics[width=0.49\textwidth]{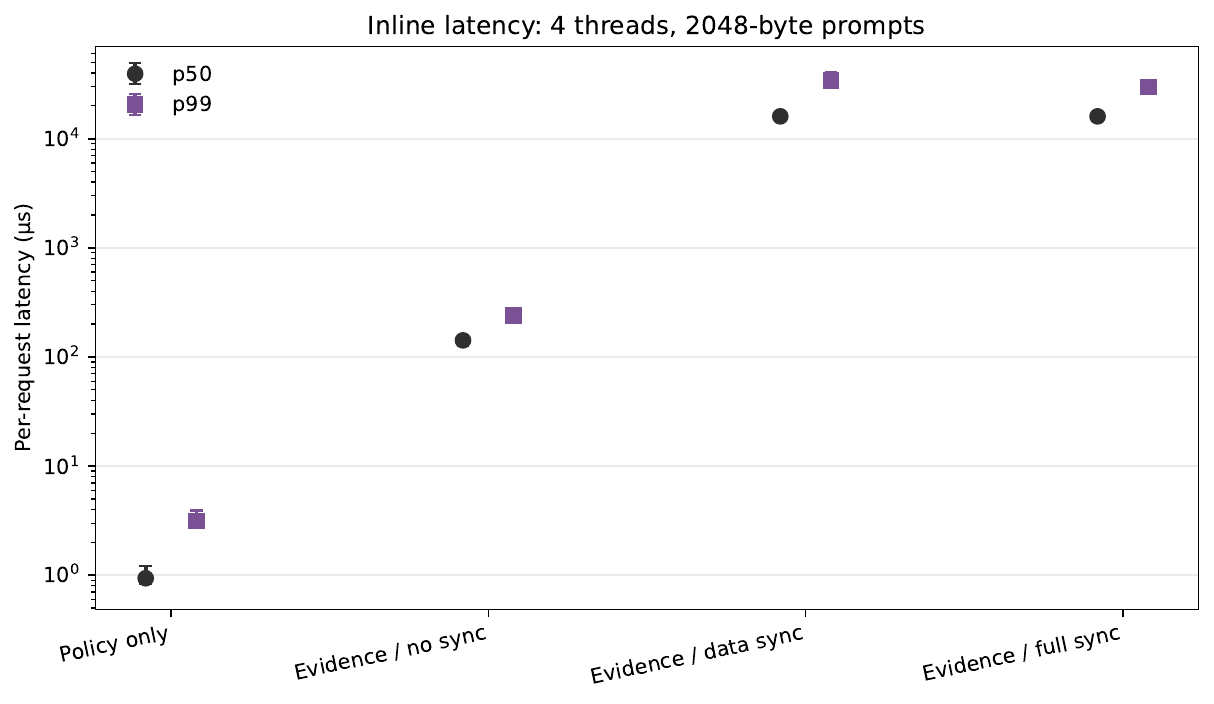}\hfill
  \includegraphics[width=0.49\textwidth]{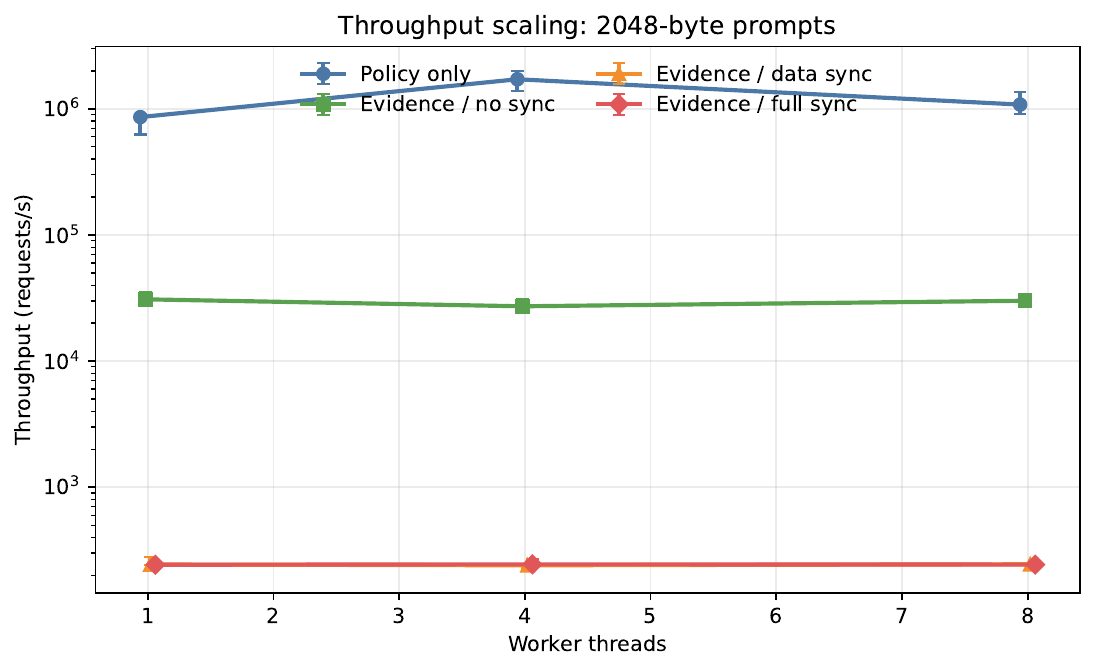}
  \caption{Inline latency and throughput from canonical raw observations. Log axes expose the orders-of-magnitude separation between policy evaluation, buffered evidence, and per-record synchronization.}
  \label{fig:inline-results}
\end{figure*}

\paragraph{RQ3: epoch construction scales with batch size.}
At \RGEpochRecords{} records, building and signing an epoch takes \RGEpochSealUS~$\mu$s (97.0~ms). Inclusion-proof verification takes \RGEpochProofUS~$\mu$s and Ed25519 statement verification takes \RGEpochSignatureUS~$\mu$s. Seal time grows from 0.116~ms at 100 records to 97.0~ms at 100,000 records, approximately tracking the amount of committed data. Proof and signature verification remain microsecond-scale over the tested sizes.

\paragraph{RQ4: recovery is near-linear in retained records.}
Opening and validating \RGRecoveryRecords{} records takes \RGRecoveryOpenMS~ms; reading and globally sorting them takes an additional \RGRecoveryReadSortMS~ms. The combined measured recovery path is 961.0~ms. The corresponding open times are 6.8~ms at 1,000 records and 66.2~ms at 10,000 records. These results support bounded single-host recovery claims, not instant restart or multi-host availability.

\begin{figure*}[t]
  \centering
  \includegraphics[width=0.49\textwidth]{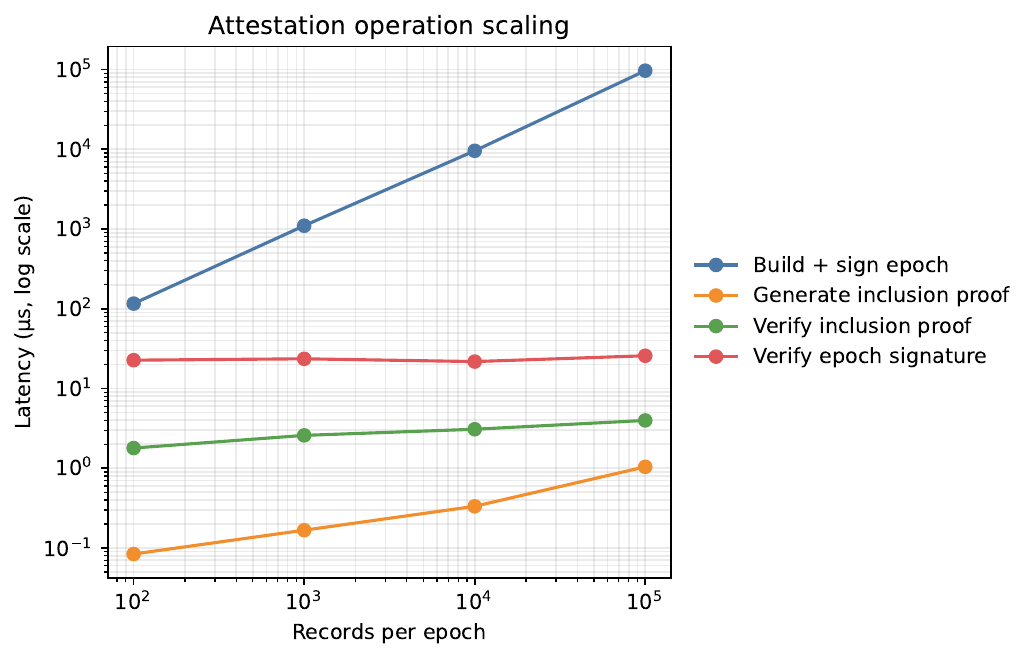}\hfill
  \includegraphics[width=0.49\textwidth]{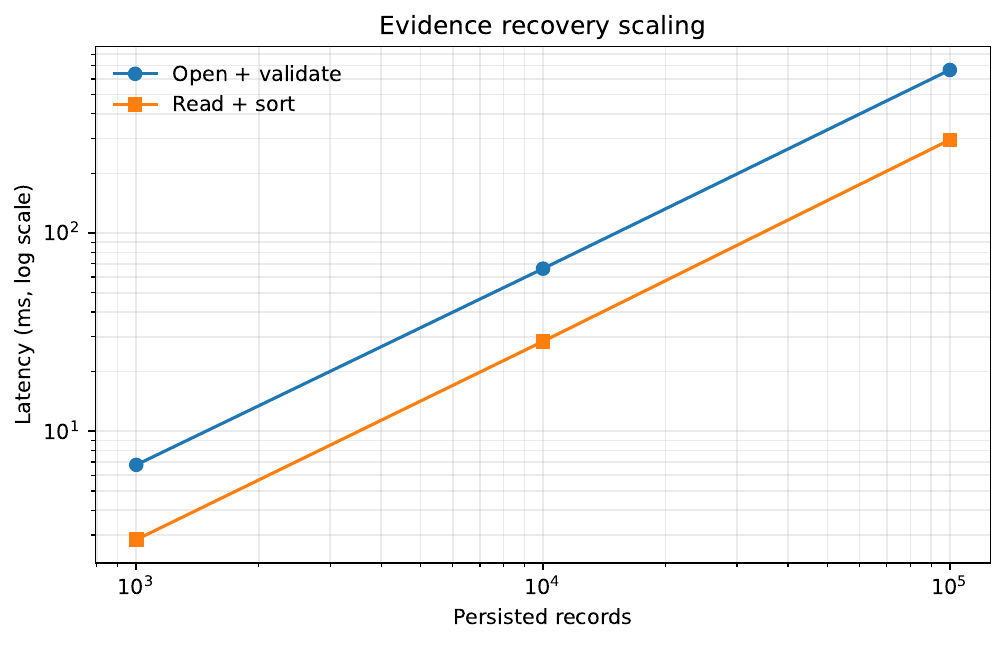}
  \caption{Canonical epoch-operation and recovery scaling. Every plotted value is derived from manifest-verified raw CSV files.}
  \label{fig:batch-results}
\end{figure*}

\section{Related Work}
\paragraph{Secure and tamper-evident logging.}
Forward-secure audit logs predate current AI systems~\cite{schneier1999secure}. Certificate Transparency standardizes Merkle commitments, signed tree heads, and consistency mechanisms for a public ecosystem~\cite{rfc9162}. Nitro shows that high-performance tamper-evident logging remains an active systems problem and uses co-design to reduce audit cost~\cite{zhao2025nitro}. These works preclude claims that Merkle inclusion, signed roots, or tamper-evident logging are novel here. \system{} instead studies an AI-specific commit interface: policy identity, decision evidence, explicit per-request durability, and the cost of that composition.

\paragraph{Transparency and trust operations.}
Sigstore demonstrates that signatures become operationally meaningful only with identity, trust roots, transparency services, and monitorable public state~\cite{newman2022sigstore}. \system{} implements externally anchored verification keys and predecessor-linked epochs, but no public service, gossip, witness, or revocation protocol. It should therefore be read as a local evidence mechanism, not a transparency ecosystem.

\paragraph{AI accountability infrastructure.}
Ojewale et al. identify infrastructure gaps beyond one-shot AI evaluations~\cite{ojewale2025infrastructure}; later work proposes lifecycle-wide LLM audit trails that join technical and governance events~\cite{ojewale2026audittrails}. Aegis proposes cryptographic runtime governance and immutable logging for autonomous agents~\cite{mazzocchetti2026aegis}. SIGIL binds audits to loaded LLM skill artifacts through an on-chain registry~\cite{shen2026sigil}, while property attestation targets claims about training-data distributions~\cite{duddu2024attesting}. These systems occupy broader lifecycle, governance, supply-chain, and property-proof spaces. \system{} is narrower: it implements and measures the acknowledgment-to-recovery path for one recorded policy decision.

\paragraph{Regulatory context.}
The EU AI Act's Article 12 directly addresses automatic record keeping; Article 14 concerns human oversight, including competence, authority, interpretation, intervention, override, and automation-bias risks~\cite{eu_ai_act}. Logging can support an oversight process but does not satisfy those obligations by itself. The NIST AI RMF likewise provides governance context, not a conformance certificate~\cite{nistairmf}. We use these sources as motivation and avoid legal-compliance claims.

\section{Limitations}
The prototype evaluates one deterministic regex policy fixture, not an arbitrary policy language or model inference stack. The benchmark is closed-loop, single-host, and component-level. It does not measure network service latency, offered-load saturation, multi-host availability, energy, key-management operations, or production SLOs.

The writer lease and commit mutex intentionally serialize durable commit ordering. Shards distribute storage files but do not create parallel sequence authorities. The system does not authenticate caller-supplied model identity, user identity, or timestamps. It records commitments to those assertions.

Checksums alone do not resist a privileged operator who rewrites complete records. Signed receipts and observed epochs expose some later omissions, but a signer can suppress unsealed records or fork views unless clients compare heads or use an external witness. Whole-directory rollback to an older valid prefix is out of scope. Key rotation, revocation, hardware isolation, and public checkpoint publication remain future work.

Finally, no cryptographic relation proves policy or model execution. The exact-source policy binding prevents accidental or caller-driven mislabeling inside the trusted implementation; it does not protect against replacement of the implementation itself.

\section{Conclusion}
Runtime accountability begins at the acknowledgment boundary. A system that returns before durable evidence exists must say so; a system that claims durability must pay and measure the corresponding storage cost. \system{} turns that boundary into a signed interface, couples it to strict recovery invariants, and extends committed records into independently verifiable signed epochs. The result is intentionally smaller than the architecture claimed by the version-of-record paper, but every surviving claim maps to executable code, tests, or source-hashed evidence. Stronger guarantees require additional mechanisms rather than stronger adjectives: external witnesses for fork detection, trusted execution or proof systems for evaluator integrity, and operational key lifecycle for long-lived trust.

\begingroup
\footnotesize
\bibliographystyle{abbrv}
\bibliography{references}
\endgroup

\end{document}